\documentclass[intlimits,twoside,a4paper]{article}

\usepackage{amsmath,amssymb}
\usepackage{graphicx}

\usepackage[T2A]{fontenc}
\usepackage[cp1251]{inputenc}

\usepackage{color}

\usepackage[eqsecnum]{cmpj2}

\issue{2016}{19}{1}{13803}
\doinumber{10.5488/CMP.19.13803}

\title[Proteins: Fractal surfaces in solutions]%
{Proteins in solution: Fractal surfaces in solutions%
}

\author[R. Tscheliessnig, L. Pusztai]{R. Tscheliessnig\refaddr{label1},
L. Pusztai\refaddr{label2}}
\addresses{
\addr{label1} Austrian Centre for Industrial Biotechnology (ACIB), Muthgasse 11, A-1190 Wien, Austria
\addr{label2} Wigner Research Centre for Physics, Hungarian Academy of Sciences, \\ Konkoly Thege \'ut. 29-33,
H-1121, Budapest, Hungary
}

\date{Received November 23, 2015, in final form December 24, 2015}

\begin{document}

\maketitle

\begin{abstract}
The concept of the surface of a protein in solution, as well of the interface between protein and 'bulk solution', is introduced. The experimental technique of small angle X-ray and neutron scattering is introduced and described briefly. Molecular dynamics simulation, as an appropriate computational tool for studying the hydration shell of proteins, is also discussed. The concept of protein surfaces with fractal dimensions is elaborated. We finish by exposing an experimental (using small angle X-ray scattering) and a computer simulation case study, which are meant as demonstrations of the possibilities we have at hand for investigating the delicate interfaces that connect (and divide) protein molecules and the neighboring electrolyte solution.
\pacs 87.10.Tf, 87.14.E-, 87.15.A-, 87.15.N-
\end{abstract}

\section{Introduction}
The appearance of aqueous solutions of even large proteins is, in many cases, similar to that of dilute solutions of simple salts: the liquid may be completely transparent, even though the size of solute molecules may be two orders of magnitude larger than that of the solvent particles (i.e., dozens of nanometers). This is made possible by strong interactions between the charged `surface' of a protein and the dipolar solvent molecules that surround a large particle; sometimes even tiny changes of the conditions (of e.g., composition, temperature) can alter the situation completely and make protein molecules aggregate and precipitate (see, e.g., reference~\cite{role-charges-a,role-charges-b}).

The `surface' of protein molecules in aqueous solutions may be considered as being defined by the hydration sphere of a macromolecule. The natural tool for studying the hydration structure, within distances of a few~{\AA}, would be wide angle X-ray (and/or neutron) scattering~--- just as it is routinely done for solutions of simple salts (see, e.g., reference~\cite{Mile}). However, due to a large number of components in a solution (water, protein, stabilizers), as well as due to the complicated internal structure and relatively low molar concentration of the protein, this route has not been very frequently chosen; examples of such studies are references~\cite{WAXS-1,WAXS-2}.

Perhaps surprisingly, it is the microscopic dynamics of the hydration sphere that has been more extensively studied than the static structure: this can be readily understood by considering that most of the dynamical studies are based on examining the dynamics of water molecules only. NMR spectroscopy~\cite{NMR-1,NMR-2}, dielectric relaxation spectroscopy~\cite{diel-1,diel-2}, as well inelastic neutron scattering~\cite{Neutron-1a,Neutron-1b} have all been applied for the purpose. More recently, terahertz (THz) spectroscopy has been used for tracking changes of the broadly defined hydration layer, up to a thickness of about 1~nm~\cite{tera-1,tera-2}.

In the pursuit of revealing the surface of a protein molecule in solutions, small angle scattering (SAS)~\cite{SAS-1,SAS-2,SAS-3} is our chosen experimental method for the present report. SAS provides a (or arguably, the only) viable experimental possibility for studying the shape of a biomolecule in solution, as it has been exemplified in  references~\cite{SAS-3,SAS-protein-a,SAS-protein-b}. Unfortunately, the interpretation of SAS data is far from being straightforward: this issue is considered in detail later in this work (see below).

In any case, to make the surface of a biomolecule `visible', one needs to possess a high (most preferably, atomic) resolution picture of the molecule in solution. No experimental technique is capable of providing such pictures so far: for this reason, we must turn to computer simulations, such as the molecular dynamics (MD) method~\cite{Allen}. Proteins and their solutions have been targeted by MD for quite some time, due to the pioneering works of Karplus and co-workers (see, e.g., reference~\cite{MD-1}). The MD methodology will also be made use of extensively in the present work; more details will be provided in due course.

One way of defining the surface of a protein is to evaluate the `solvent inaccessible' volume of the biomolecule. In the cube method \cite{contour-1}, the biomolecule is placed in a parallelepiped-shape box which is subdivided into small cubes with edges of 0.5--1.5~{\AA}. The boundary of the biomolecule is determined by examining whether each cube belongs to the biomolecule or to the solvent (see, e.g., reference~\cite{contour-2}). A more complicated method is to calculate the `electron envelope' of the macromolecule: an algorithm for this is implemented in the program CRYSOL17 \cite{contour-3}.

In general, due to a large variety of the ways the beta sheets and alpha helices are put into sequences in protein molecules, the surface of such molecules is rather complicated. In the present contribution, we consider that in general (or at least, in a large number of cases) the boundary of a protein molecule may have a fractal dimension. We pursue this idea by presenting theoretical and experimental arguments; we finish with providing computer simulation results based on simple concepts.

\section{Scattering from fractal surfaces}

What is a surface in terms of scattering theories? Small angle scattering may provide information on surface areas that are larger and more uniform than that of a biomolecule (see references~\cite{SAS-1,SAS-2,SAS-3}); we must, therefore, take a more indirect way. In fact, connections between the measured intensity and fractal (or `rough') surfaces have already been sought for \cite{sas-fract-1,sas-fract-2,sas-fract-3}; note that these investigations have not considered protein surfaces directly.

\begin{figure}[!b]
\centerline{
\includegraphics[scale=.85]{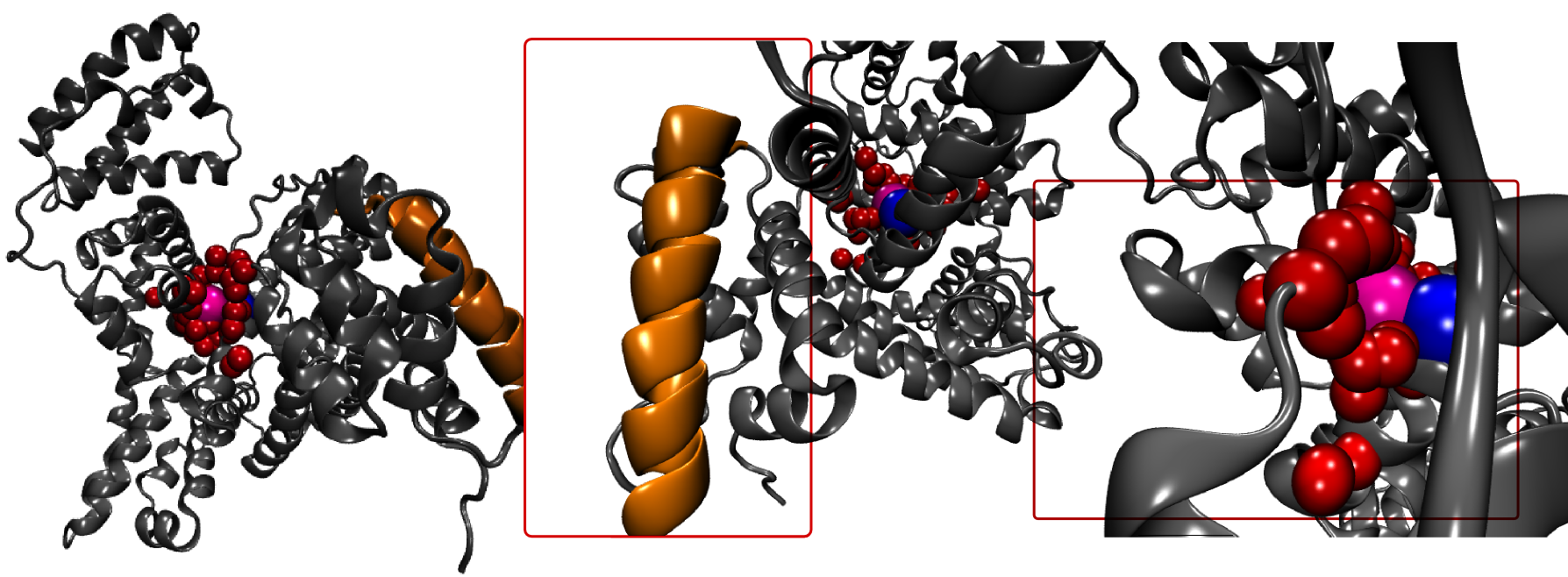}
}
\caption{\label{fig:01} {(Color online) The point of reference.} Left-hand panel: Any protein is a complex structure of scattering sites. The issue to decide upon: which is the one we refer to? What are then reaction coordinates? Red spheres mark the sites for which the Fourier transform of pair and  radial distributions provides comparable results. Their centroid is colored magenta, while blue spheres mark the centroid of all of the protein sites. (From left to right: the important part considered is gradually enlarged.)}
\end{figure}

While the `reaction coordinate', i.e., the location of a site of importance (e.g., of a scattering site) within the investigated volume  in a slit pore seems obvious, as it follows from  the symmetry of the pore, it  is  a complex task to determine if we deal with soft matter, e.g., proteins. Let us take a rather simple protein: it will be  formed by alpha helical domains (a typical one is indicated orange in figure~\ref{fig:01}) and joined by random coils. For the present considerations, we have chosen a  well-known globular protein, selected out of thousands of possibilities: Bovine serum albumin (BSA) (for its crystalline structure see reference~\cite{BSA-structure}).  We determine its point of reference (in other words, the `origin' of the system). This is a crucial step because  it will mathematically determine  what we term a {\it fractal} surface.

We assume scattering sites in the vicinity of, or indeed, within amino acids. We compute their centroid and from their relative distance we compute the pair densities. The chosen system lacks any symmetry and that is why we use equation (\ref{eq.02}) to determine the point of reference and compute, with respect to it, the radial density function.

First, we  sketch a mathematical methodology to access structural information from small angle and neutron scattering data; this information will be related to the issue of the surface of a protein. We link the distribution of scattering sites to the definition of $\alpha$ stable distributions \cite{zolotarev, mandelbrot}.

We assume that $\{X_i\}$ and $\{X_j\}$ are random variables. Here, they are distances of scattering sites,  with respect to sites $i$ or site $j$ of those variables, and  they are distributed according to a particular probability density $\phi(\zeta)$. The distribution is called stable if the probability density
$p(\zeta)$ of any linear combination $Y=\lambda_1X_1+\lambda_2X_2$ then,
\begin{equation}
\gamma(\zeta) = \lambda_1 \lambda_2 \int_0^\infty \rd\zeta' \phi\big((\zeta-\zeta')\lambda_1\big) \phi(\zeta'\lambda_2) .\label{eq.01}
\end{equation}

The distributions coincide subject to rescaling, i.e.,
 \begin{equation}
\langle \mathcal{F}(\gamma(\zeta))[Q] \rangle= \langle |\mathcal{F}(\phi(\lambda\, \zeta))[Q]|^2 \rangle.\label{eq.02}
\end{equation}

It is a significant extension to the formulation of Kotlarchyk's work~\cite{Rupert-Kotlarchyk}, as it includes  scaling
to relate the pair density  of scattering sites of a protein with their radial distribution.

Kotlarchyk relates the scattering intensity, $I(Q)$, to the protein form factor $P(Q)$ and the protein-protein structure factor $S(Q)$ by
\begin{equation}
I(Q)=P(Q )\left\{1 +\beta(Q)[S(Q)-1]\right\}\label{eq.03},
\end{equation}
wherein the term $\beta(Q)=P^\star(Q )/P(Q )$ takes into account the possible anisotropic form of a protein. In the previous paper~\cite{Rupert-own-1}   we introduced the fractal pendant to Debye's formula:
\begin{equation}
\mathcal{J}_D[Q\,\zeta_b ]=\frac{J_{D/2-1} (Q\,\zeta_b)}{(Q \zeta_b)^{D/2-1}}.\label{eq.04}
 \end{equation}

We did give clear evidence~\cite{Rupert-own-1} that the fractal dimension, $D$, may be related to the Debye screening length, and that it is not necessarily $D=3$.

We rewrite the scattering intensity:
 \begin{eqnarray}
&& I(Q) = \mathcal{F}_D(\gamma(\zeta_R))[Q]
= \int_0^\infty \rd \zeta_R \zeta_R^{D-1} \gamma(\zeta_R)\mathcal{J}_D[Q\,\zeta_b ]\label{eq.05},
\end{eqnarray}
as a function of the pair density  of the protein scattering sites $\gamma(\zeta_R)$. The pair density  is a function of the relative distances, $\zeta_R$, between individual protein scattering sites. The protein form factor
 \begin{eqnarray}
&& P(Q)=\langle |F(Q)|^2\rangle =\mathcal{F}^2_{D}(\phi(\zeta_b))[Q]
= \int_0^\infty \rd \zeta_b \zeta_b^{D-1} \phi(\zeta_b) \mathcal{J}_D[Q\,\zeta_b ]^2,
\end{eqnarray}
however, is the Fourier transform of the radial probability density. It is a function of $\zeta_b$, the distance with respect to an arbitrary site, within the protein. Commonly, the center of mass of the protein centroid is chosen.

In order to compute the protein anisotropy, $\beta(Q)$, we need
\begin{eqnarray}
&&P^\star(Q)=|\langle F(Q)\rangle|^2 =\mathcal{F}_{D}(\phi(\zeta_b))[Q]^2
= \left(\int_0^\infty \rd \zeta_b \zeta_b^{D-1} \phi(\zeta_b)\mathcal{J}_D[Q\,\zeta_b ]\right)^2.
\end{eqnarray}

We shall not explore a detailed deduction but draft the essence, and provide motivation from the observation of scalability by wet lab and computer experiments
 \begin{eqnarray}
&&I(Q) =\mathcal{F}_D(\gamma(\zeta_R))[Q]
= \int_0^\infty \rd \zeta_R \zeta_R^{D/\lambda-1} \gamma(\zeta_R)\mathcal{J}_{D/\lambda}[Q\,\zeta_b ]\label{eq.06}.
\end{eqnarray}

Due to the scaling capability and alpha stability of the pair density  and radial probability density,  we are allowed to introduce $\zeta_b=\lambda\zeta_R$ and rewrite the scattering intensity in terms of the protein form factor:
\begin{eqnarray}
 P(Q) &\propto& \int_0^\infty \rd\zeta_b\zeta_b^{D-1} \phi(\zeta_b/\lambda) \mathcal{J}_D[Q\,\zeta_b]^2
 = \lambda^D \int_0^\infty \rd\zeta_R  {\zeta_R}^{D-1}\phi(\zeta_R) \mathcal{J}_D[Q\,\lambda \zeta_R]^2\nonumber\\
 &\propto&    \int_0^\infty \rd\zeta_R  {\zeta_R}^{D/\lambda-1}\phi(\zeta_R) \mathcal{J}_{D/\lambda}[Q \zeta_R]^2.
 \end{eqnarray}

The above is a set of equations that we term as `fractal scattering theory'.

\section{Small angle neutron and X-ray scattering from biological soft matter}

\begin{figure}[!b]
\centerline{
\includegraphics[width=0.99\textwidth]{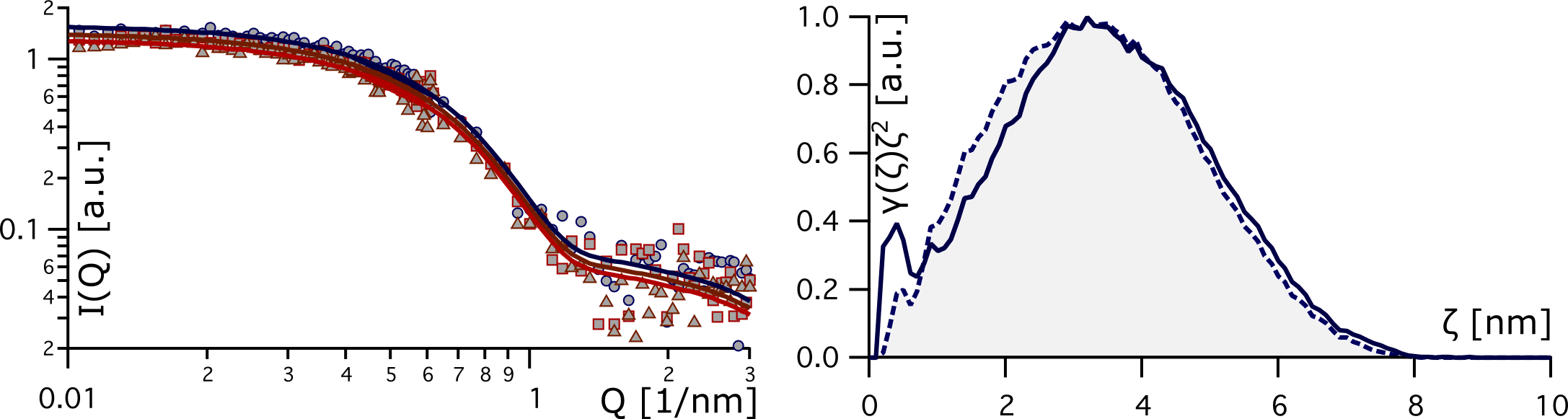}
}
\caption{\label{fig:02} (Color online) {SANS data of BSA at different salt concentrations}. Left-hand panel: SANS data of BSA dissolved in three different environments which contained zero ammonium sulfate (blue squares), 0.7~mol/kg (i.e., per kg of solution) ammonium sulfate (red circles), and 1.2~mol/kg (NH$_4$)$_2$SO$_4$ (brown triangles).
Solid lines give fits of experimental data. Right-hand panel: Dashed  blue line with grey markers indicates the pair density computed from a crystallographic model
of BSA, whereas solid blue line marks the pair density that corresponds to fits shown in the left-hand panel. }
\end{figure}

In this section we briefly discuss the origin of the fractal dimension $D$.

Small angle neutron and small angle X-ray scattering data were collected  to obtain structural information for BSA (concentration: 5~mg/ml) in three different aqueous salt environments (i.e., in different electrolyte solutions). The data are displayed in figure~\ref{fig:02}. These three different environments contained zero ammonium sulfate (state~1), 0.7~mol/kg ammonium sulfate (state~2), and 1.2~mol/kg (state~3). The pH of the solutions is very close to neutral  (just below~7) and these electrolyte concentrations are far below the salting-out limit of the protein. Detailed description of the experiments can be found in references~\cite{Rupert-own-1,Rupert-own-2}.

We argue that the parameter $D$ is considered to be of electrostatic origin and proportional to the salt concentrations in bulk solutions \cite{Rupert-own-2}.  In figure~\ref{fig:02} (right-hand panel) the pair density  function of the initial crystallographic model is shown
as dashed blue line with grey marker. The fits (solid lines, left-hand panel) were obtained by calculating the pair density function (solid blue
 line, right-hand panel) and scattering intensity using equation (\ref{eq.04}), with $D=3$. Clearly, irrespective of the ion concentrations, we do not see significant changes in terms of the electronic contrast.

\begin{figure}
\centerline{
\includegraphics[width=0.99\textwidth]{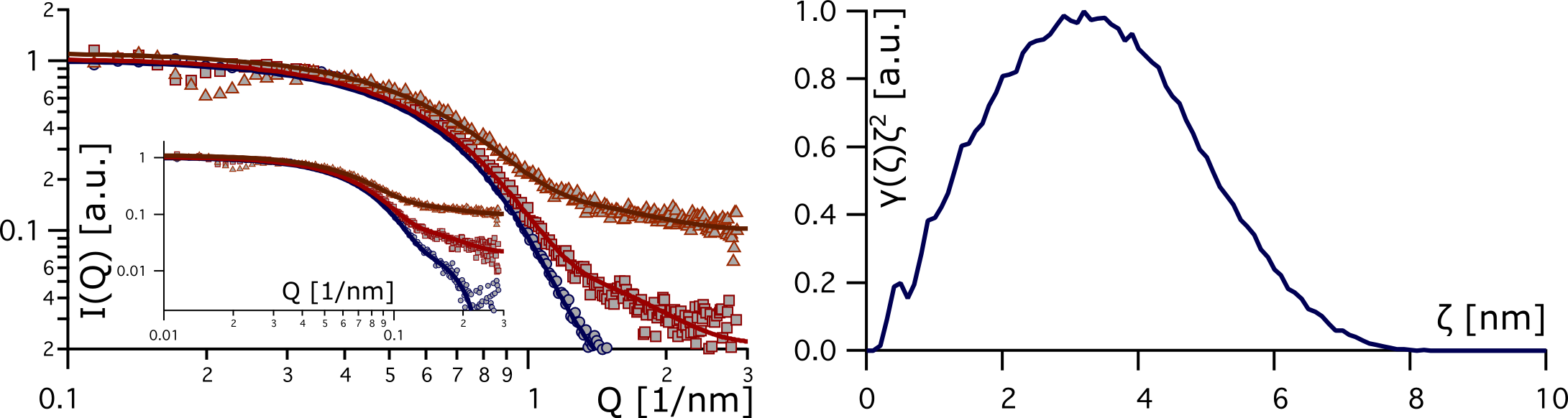}
}
\caption{\label{fig:03} {(Color online) SAXS data of BSA at different salt concentrations}. Left-hand panel: SAXS data of BSA dissolved in three different environments which contained zero ammonium sulfate (blue squares), 0.7~mol/kg ammonium sulfate (red circles), and 1.2 mol/kg (brown triangles). Solid lines give fits of experimental data, for $D = 3.2$, $D = 2.9$ and $D = 2.5$.
Right-hand panel: Solid blue line indicates the pair density computed from a crystallographic model of BSA. All data presented in the left-hand
panel were computed from it for different fractal dimensions.}
\end{figure}

The SANS measurements were complemented by SAXS measurements for identical solutions. SAXS data are presented  in figure~\ref{fig:03}. Note the discrepancy between the results of the two experimental techniques. Though the systems are identical, their scattering intensities $I(Q)$ differ.

Typically, for small angle X-ray scattering, one is tempted to interpret the changes in $I(Q)$ by the changes in their individual pair density distributions, and then,  consequently, argue the changes in the protein conformation. However, this line of arguments is not supported by small angle neutron scattering data.

In what follows, we interpret the data differently: we leave the pair correlation untouched and change the parameter $D$, which may be interpreted as a fractal Dimension.  Note that the fits of experimental data have been achieved by changing $D = 3.2$, $D = 2.9$ and $D = 2.5$ without changing the protein conformation. We left the density distribution of the protein untouched.
The linear relation of $D$ to the ionic strength of the particular solution is obvious. The higher the salt concentration the lower $D$ should be put.

For a quick exploration of the effects of varying $D$, we use a computational approach, using molecular dynamics simulations (see next section).

\section{The (fractal) surface of biomolecules: demonstration via computer simulation}

Having defined three different quantities, i.e., the form factor, the structure factor and the anisotropic factor [$\beta(Q)$], it is time to explore these and put them in relation to computational approaches, such as density functional theory~\cite{Stefan-paper-1}. Therefore, we set up three systems. We discuss two of them qualitatively, whereas the third one we explore in detail. Since many theoretical systems, especially in the density functional theory, deal with slit pores~\cite{Stefan-paper-1},  we shall start with these.

From a mathematical point of view it is difficult to compute the pair distribution of an infinite planar slit pore numerically, as one would need to  compute the pairwise densities over all sites of a slit pore. The sum, or moments of the sum, would not necessarily converge: one might think of particle interactions that produce in plane pair densities that we may consider $\alpha$ stable. The common way out is to measure and compute  density distributions perpendicular to the surface.

Let us switch to spherical coordinates: we do so for different reasons. They seem mathematically easier as well as they are very frequently applicable in soft matter as many a system investigated is of spherical symmetry. In fact, it may be the experiment as well  that imposes spherical symmetry to the measured data, just as small angle X-ray and neutron scattering certainly do (see the previous section).

Let us rethink the planar slit pore to be an infinite spherical one. Then, we have to consider the point of reference, in order to define a reaction coordinate, $\zeta$.  For planar slit pores, its particular symmetry suggests to place the point of reference in the center of the slit pore. This may also be used for finite and infinite spherical slit pores. Now, a spherical slit pore will consist of two concentric spheres. The inner shell has a radius of $r_\infty$ while the outer one, a radius of $r_\infty+\Delta$. We define a {\it radial density distribution} by exploiting the shift property of the Fourier transform:
\begin{equation}
 \langle |\mathcal{F}(\phi(\lambda\, (r_\infty+\zeta)))[Q]|^2 \rangle = \langle |\mathcal{F}(\phi(\lambda\, \zeta))[Q]|^2 \rangle.
\end{equation}

We hereby reinterpret the planar slit pore to be an infinite spherical slit pore. We shift the point of the origin next to one planar surface since it would be numerically cumbersome to compute the pair distribution of an infinite spherical slit pore, but by the use of  equation (\ref{eq.02}). Next, we drop the inner spherical surface and replace it by a `protein'. We use a simple Lennard Jones (LJ) model.  We used the molecular dynamics package LAMMPS \cite{lammps}. All parameters listed in the subsequent paragraphs are reduced to the wall LJ parameters. To save computational time, we rescale the protein by a factor of five.

\begin{figure}[!b]
\centerline{
\includegraphics[width=0.76\textwidth]{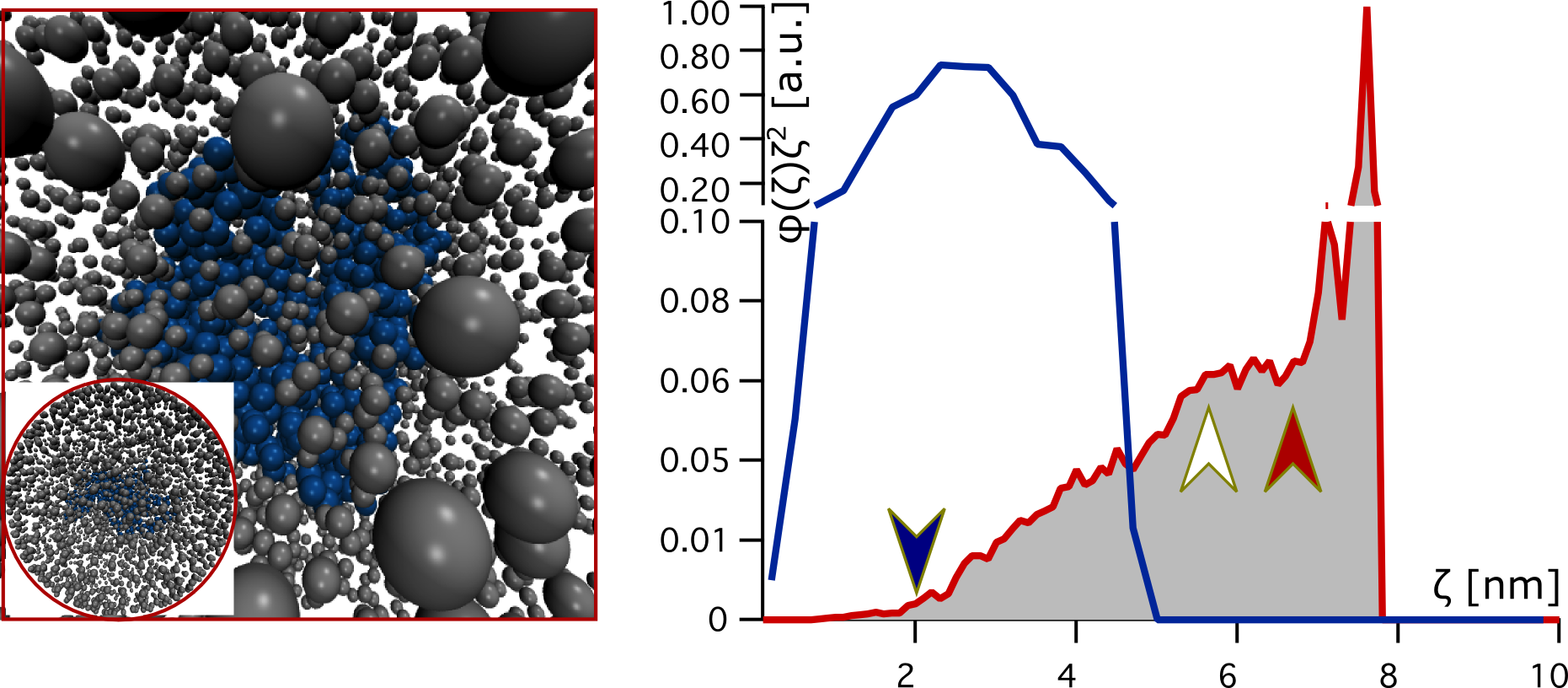}
}
\caption{\label{fig04} {(Color online) MD simulation of a protein dissolved in a LJ liquid.} Left-hand panel: a protein (blue beads) is dissolved in a LJ liquid (grey beads). The simulation box is not periodic: both types of particles are enclosed in a spherical wall. Right-hand panel: the radial density, $\phi(\zeta)$ is displayed for the protein (blue) and the LJ liquid (red line with grey markers). We  distinguish the protein, the protein surface and the LJ liquid bulk. }
\end{figure}

We compute the centroids of each amino acid and replace these by LJ sites. `Pair styles', i.e., specific parameters for the particular pairs of sites (for details, see the LAMMPS Manual \cite{lammps-manual}) between protein and liquid were put to  $\epsilon=0.1$ and $\sigma=2.5$. The protein is  positioned in its appropriate center, as computed from (\ref{eq.02}). For simplicity, we fix the protein ``amino acids'' by springs to the centroids. The spring constant that kept sites of the protein was put to $k = 10$. This value was chosen so that the liquid may slightly penetrate the protein. Interactions within the protein were turned off. The protein is dissolved in a LJ liquid.  For liquid-liquid interactions, we constructed a hybrid potential by superposing two pair styles, a {\it lj/soft/cut} and a {\it gauss/cut}. LJ parameters for liquid-liquid interactions were set to  $\epsilon = 0.05$ and $\sigma= 1.5$. A repelling Gaussian potential was added  to the liquid-liquid interactions, whose amplitude was set to 0.05. A repelling distance of $\zeta =  1.0$ and a variance of 1 were used. The liquid  comprised 3553 sites.

The construction of wall and liquid is enclosed in a spherical wall of a diameter $\zeta_\text{d} = 32$. It is a LJ wall of type  {\it wall/lj93} (according to LAMMPS terminology) and parameterized as $\epsilon = 1.0$ and $\sigma = 1.0$.

We performed simple NVE simulations and initially gave all sites to a velocity of 3. After equilibrations of 500 steps, we performed simulations of 5000 steps. The system was reduced to a configuration as shown in figure~\ref{fig04}.
In the right-hand panel of figure~\ref{fig04} we find the radial distribution of  the protein (solid blue line) and the radial distribution of the LJ liquid
(solid red line with grey markers).
Both were normalized to their maximum value. We rescaled these results to run them comparable to the experimental data.  Arrows  in  figure~\ref{fig04} right-hand panel mark three regions. The reaction coordinate up to the blue arrow is termed protein. We attribute the linear regime (in-between blue and white arrow) to the protein surface, whereas the planar regime (in-between white and red arrow) is attributed to the LJ liquid bulk. The corrugation in the radial density of the LJ liquid around 8~nm proves the liquid-like state of it.

\begin{figure}[!t]
\centerline{
\includegraphics[width=0.96\textwidth]{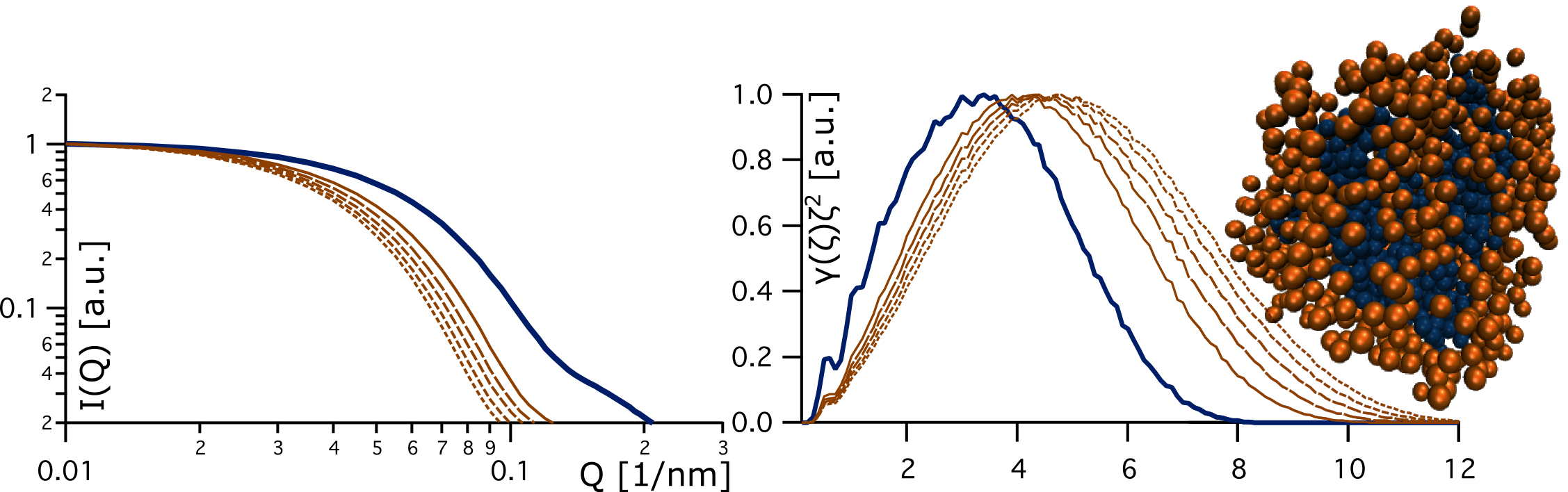}
}
\caption{\label{fig05}(Color online) Self similar SAXS signals. Left-hand panel: SAXS
 profiles computed from pair density distributions. The blue solid line
 mimics the scattering profile computed from the protein crystallographic
 model. There was no background added or subtracted. The orange (solid and
 dashed) lines give scattering profiles computed from the protein and from
 particles from the LJ liquid in the proximity of the protein. Right-hand
 panel: The blue solid line gives the pair density distribution for the LJ
 protein. The full and dashed orange lines indicate pair density
 distributions computed from the LJ protein and particles from the LJ liquid
 that are in the proximity of the protein. They scale invariantly.}
\end{figure}

In figure~\ref{fig05},  scattering profiles for protein plus protein surfaces of different thicknesses are displayed. All these complexes are in the linear regime shown by the insert of figure~\ref{fig04}. While the blue line refers to
the hypothetical scattering profile of a blank protein, the orange lines refer to scattering
profiles and pair densities of the protein embedded in LJ liquid of different thickness. The pair densities are self-similar. The larger is the construct, then  the corresponding scattering profile is found more to the left.

In the structure model in figure~\ref{fig05}  we discriminate the protein (blue beads) from the protein surface (i.e., the `hydration shell' of the protein, orange beads).
The protein surface was determined as follows. For each amino acid we computed ten closest LJ sites. These form the protein surface.
Clearly, we do see areas of low density of LJ sites in the protein surface surrounded by areas of high density of LJ sites.

{It is evident that within the hydration shell, the local density of LJ sites differ. Their distribution is (though influenced by the parameters chosen) altogether a consequence of the protein morphology. It is a key difference from planar surfaces, where we expect a homogeneous distribution perpendicular to the surface.}

Another difference is the linearity of the hydration shell, while the spherical surface already enforces a layered structure. This seems to suggest that protein fractal morphology extends the Henry regime to higher bulk densities~--- a conjecture that needs clarification in the future.

\vspace{-1mm}
\section{Summary and outlook}
\vspace{-1mm}

In this work we provide a (somewhat limited) collection of mathematical formulae that may be useful to link theoretical findings of classical density functional theory to experimental results derived from scattering techniques, such as small angle neutron and small angle X-ray scattering. We discuss the necessity of these and their fractal flavour. Though we lack a detailed mathematical discussion of the possible physical origin, we have experimental evidence that may be found in the electrostatics of the system investigated. We compare experimental data from small angle neutron scattering to the data of small angle X-ray scattering. While neutron scattering data do not change upon different salt concentrations, small angle X-ray data do. These changes in the scattering data can be explained by a fractal dimension, which is of electrostatic origin. We performed molecular dynamics simulation and presented a structure model. We distinguish protein from protein surface and find scale invariance for both.

\vspace{-1mm}
\section*{Acknowledgements}
\vspace{-1mm}

ACIB  is supported by the Federal Ministry of Economy, Family and Youth (BMWFJ), the Federal Ministry of Traffic,
Innovation and Technology (BMVIT), the Styrian Business Promotion Agency SFG, the Standortagentur Tirol
and ZIT-Technology Agency of the City of Vienna through the COMET-Funding Program managed by the Austrian Research Promotion Agency FFG.  LP aknowledges financial support from the National Research, Development and Innovation Office of Hungary (NKFIH), grant no.~SNN~116198.

\vspace{-6mm}

\ukrainianpart

\title{Протеїни в розчині: фрактальні поверхні у розчинах}

\author{Р. Челєссніг\refaddr{label1}, Л. Пустаї\refaddr{label2}}
\addresses{
\addr{label1} Австрійський центр промислової біотехнології (ACIB),  A-1190 Відень, Австрія
\addr{label2} Вігнерівський дослідницький центр з фізики, Угорська академія наук, Будапешт,
H-1121, Угорщина
}

\makeukrtitle

\begin{abstract}
Впроваджено концепцію поверхні протеїну у розчині, а також границі розділу між протеїном та ``об'ємним розчином''. Коротко описано  експериментальний метод розсіяння X-променів при малих кутах і нейтронного розсіяння. Обговорено моделювання молекулярної динаміки як обчислювального інструменту для вивчення гідратаційної оболонки протеїнів. Розроблено концепцію протеїнових поверхонь з фрактальним виміром. Статтю завершено описом конкретного випадку
експериментального (з використанням  розсіяння X-променів при малих кутах)  та комп'ютерного дослідження, щоб продемонструвати  наявні можливості дослідження делікатних границь розділу молекул протеїну і розчину електроліту.

\keywords розчин протеїну, гідратація протеїнів, поверхня протеїну, розсіяння при малих кутах
\end{abstract}

\end{document}